\documentclass{aa}
\usepackage{graphicx}
\begin{document}


   \title{Asymptotic Giant Branch evolution  at varying surface C/O ratio:\\ 
   effects of changes in molecular opacities}

   \author{Paola Marigo}

   \institute{Dipartimento di Astronomia, Universit\`a di Padova, 
		Vicolo dell'Osservatorio 2, 35122 Padova, Italia}

   \titlerunning{Molecular opacity effects on AGB evolution}
   
   \offprints{P. Marigo \\ \email{marigo@pd.astro.it}}

   \date{Accepted for publication in A\&A}

   \abstract{
    We investigate the effects of molecular opacities on the evolution  
    of TP-AGB stars that experience the third dredge-up,
    i.e. with surface abundances of carbon and oxygen varying 
    with time.
    To this aim, a routine is constructed to derive the molecular
    concentrations through dissociation equilibrium calculations, 
    and estimate the opacities due to H$_{2}$, H$_{2}$O, 
    OH, C$_{2}$, CN, and CO for any given density, temperature 
     and chemical composition of the gas.
    Then, synthetic TP-AGB models with dredge-up are calculated 
    by either adopting 
    the newly developed routine, or interpolating between  
    fixed opacity tables for solar chemical composition.
    The comparison between the two cases shows that
    the change in the dominant opacity sources, as the C/O ratio
    grows from below to above unity, crucially affects the evolution  
    of the effective temperature, i.e. causing a notable
    cooling of the carbon-rich models (with C/O$>1$).
    From the comparison with observational data, 
    it turns out that TP-AGB models with variable molecular 
    opacities are able to reproduce the
    observed range of effective temperatures, 
    mass-loss rates, and wind expansion velocities of C-type giants 
    in the solar neighbourhood,
    otherwise failed if assuming fixed molecular opacities 
    for solar-scaled mixtures.
    Finally, we mention other possibly important evolutionary and 
    observational effects that result from the adoption of the 
    variable opacities, such as: i) significant shortening 
    of the C-star phase due to the earlier onset of the super-wind;
    ii) consequent reduction of the carbon yields; iii) 
    reproduction of the observed range of near-infrared colours of C-stars.
   }
	
   \maketitle
%

\section{Introduction}
The spectra of cool giant stars -- with spectral types M-S-C  -- 
are dominated by totally different molecular absorption bands in the visual
and infrared (e.g. Lan\c con \& Wood 2000) depending on which,
carbon or oxygen, is the most abundant element.
For instance, oxygen-rich stars exhibit strong bands of 
TiO, VO, H$_{2}$O, whereas carbon-rich stars show large absorption features 
of C$_{2}$, CN, SiC.
These striking spectral differences between oxygen-rich and carbon-rich
stars were first quantitatively 
explained  by Russell (1934), on the basis of molecular equilibria
calculations. 
The key-point is that the binding energy of the CO molecule is so high 
that almost all atoms of the least abundant element -- C or O --
are locked to form CO, while the excess atoms of the most abundant 
element are involved in the formation of the characteristic molecular bands.
In addition, the relative abundances of carbon and oxygen 
affect the chemical composition of the predominant dust grains 
-- e.g. amorphous silicates in oxygen-rich mixtures, and  
amorphous carbon grains in carbon-rich mixtures 
(see e.g. Ivezi\'c \& Elitzur 1995; Habing 1996) --  which
condensate in thick circumstellar envelopes ejected by
AGB mass-losing stars.

These spectral differences 
in molecular blanketing and dust emissivity concur  
to create an observed sharp dichotomy
in infrared colours between 
M- and C-type stars. For instance, the near-infrared colours 
(e.g. in the $JHK$ bands) of carbon stars 
are systematically redder than those of oxygen-rich stars, as
illustrated in several photometric surveys of AGB stars, e.g. 
in the Magellanic Clouds such as in 
Frogel et al. (1990), 
Costa \& Frogel (1996), Sergei \& Weinberg (2000, the 2MASS project),  
Cioni et al. (2000, the DENIS project).
An analogous situation occurs in the far-infrared, e.g. in the 
IRAS two-colour diagram carbon-rich stars are displaced to
larger $[25-60]$ colours compared to oxygen-rich stars
(see e.g. van der Veen \& Habing 1988).

It is clear that theory should account for these 
observational features, 
also in consideration of the fact that huge 
amounts of infrared data are being released 
(e.g. the DENIS and 2MASS projects). 

Actually, on the theoretical side the situation is as follows.
The importance of molecular opacities in determining the photospheric 
properties of AGB stars has been demonstrated since long ago  
(e.g. Tsuji 1966). Depending on the surface C/O ratio,
the dominant sources of opacities 
at low temperatures are different, being essentially those of 
H$_2$O and TiO for
oxygen-rich atmospheres, and 
CN and C$_2$ for  carbon-rich atmospheres (see e.g. 
the reviews by Gustafsson  \&  J\o rgensen 1994, Gustafsson 1995).

On one side, sophisticated model atmospheres for late-type stars
have been constructed with the inclusion
of molecular opacity data  which are suitable for either M-type stars
(e.g. Brown et al. 1989; Plez 1992; Bessell et al. 1998), 
or C-type stars (e.g. Querci et al. 1974; Johnson 1982;
J\o rgensen et al. 1992; H\"ofner et al. 1998).

On the other side, in  most stellar evolution models of AGB stars
the description of low-temperature opacities is still inadequate.
In fact, the usually adopted prescriptions 
(e.g. Alexander \& Ferguson 1994) correspond
to opacity tables which are strictly valid for gas mixtures with 
solar-scaled abundances of elements heavier than helium, 
hence implying C/O$\sim 0.48$.
For a given initial metal content, the gas opacities are usually
derived by interpolating the data tables as a function of density, 
temperature and hydrogen abundance. This means that
any change in the true molecular opacities, due to e.g. variations of the
surface CNO abundances, is neglected.
The most notable drawback is that 
molecular opacities strictly valid for oxygen-rich configurations 
(with C/O$<1$) are applied even to carbon-enriched models (with C/O$>1$), 
regardless of the sharp dichotomy in the molecular equilibria
between the two cases.  

To our knowledge,
limited sets of molecular opacity tables at variable C/O ratios
are available in the literature (Alexander et al. 1983; Lucy et al. 1986), 
and few works are dedicated so far
to consistently couple molecular abundances and opacities
in evolutionary  calculations of carbon stars. 
Scalo \& Ulrich (1975) first calculate two AGB
evolutionary sequences with variable molecular opacities.
The effective temperatures of the models -- 
with increasing surface carbon abundance --  
are derived with the aid
of envelope integrations in which the molecular concentrations and
relative low-temperature opacities are let vary according to the 
current CNO abundances. It turns out that 
the transition from  C/O$<1$ to C/O$>1$ is marked by 
a significant cooling of the H-R tracks, essentially
triggered by the sudden increase of the CN abundance and opacity.   
The sensitiveness to molecular opacities of the Hayashi limits 
for carbon stars has also been investigated by Lucy et al. (1986),
in view of analysing the development of dust-driven winds 
at low effective temperatures.
Finally, it is worth mentioning the work by Bessell et al. (1989, 1991)
who present an analytical fit  -- 
as a function of the C and O abundances -- to the molecular opacity 
calculations by Alexander et al. (1983).

\begin{figure}
\resizebox{\hsize}{!}{\includegraphics{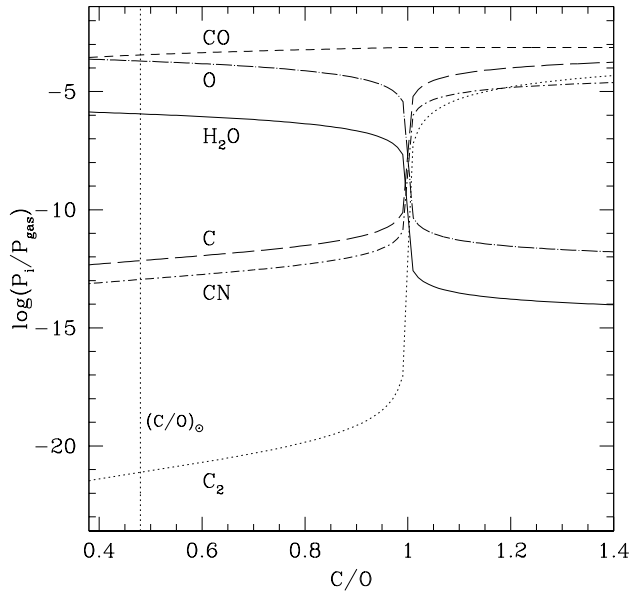}}
\caption{Relative partial pressures 
of a few atomic and molecular species as a function of the C/O ratio,
assuming a gas pressure $P_{\rm gas} = 10^3$ dyne cm$^{-2}$, and 
a temperature $T = 2500$ K.
The vertical line intercepts the predicted molecular concentrations
for a solar-like  composition with C/O$\sim 0.48$. 
Note the sharp dichotomy as C/O increases from below to above unity, 
for all species shown except the CO molecule.
In fact, due to its high binding energy, the CO molecule always locks most
atoms of the least abundant element between C and O}
\label{fig_molcovar}
\end{figure}
In this context, the present study attempts to give 
a more realistic description 
of molecular opacities and discuss their effect
 on AGB evolution 
models.
The work is organised as follows.
The adopted procedure to calculate the opacities 
is described in Sect.~\ref{sec_opac}. 
It closely resembles that developed 
by Scalo \& Ulrich (1975):
The mass absorption coefficient in any given point 
across the atmosphere is calculated  
with aid of analytical fitting relations, once the 
molecular concentrations are singled out with dissociation equilibrium 
calculations (Sect.~\ref{ssec_mol}).
Our predictions are tested by comparison with detailed  opacity
calculations  (Sect.~\ref{ssec_comp}) for solar-scaled elemental
abundances, and also examined as a function of increasing 
carbon abundance and C/O ratio   (Sect.~\ref{ssec_kco}).

\begin{figure*}
\begin{minipage}{0.72\textwidth}
\resizebox{\hsize}{!}{\includegraphics{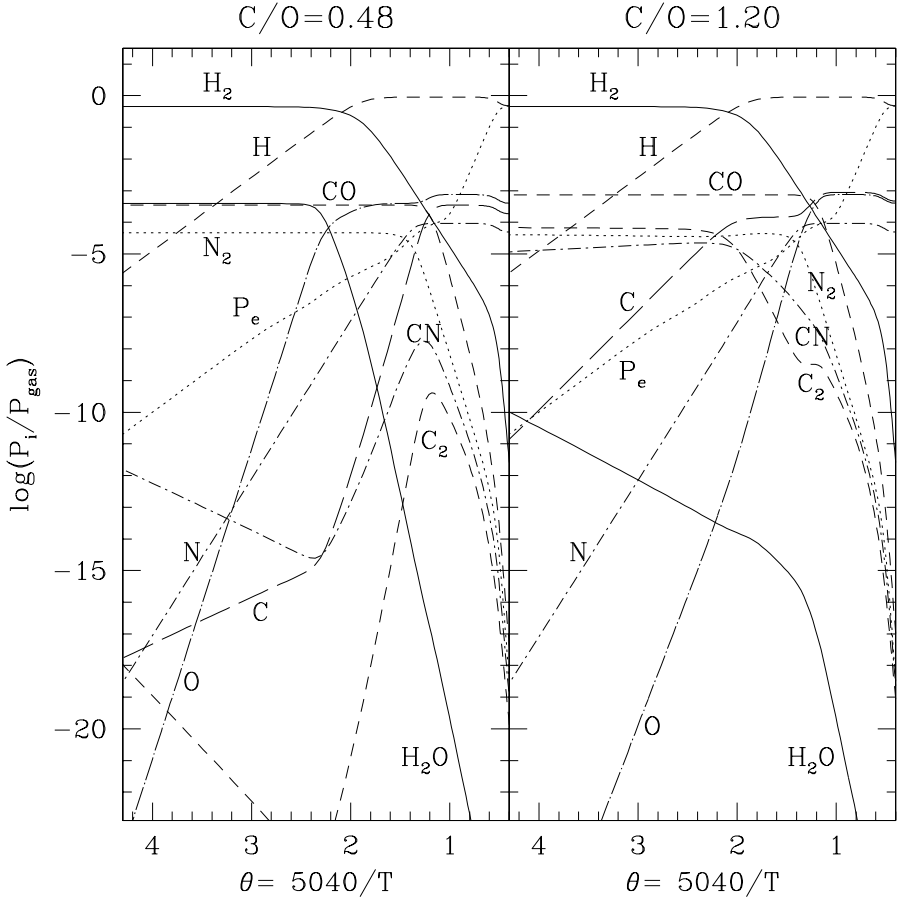}}
\end{minipage}
\hfill
\begin{minipage}{0.25\textwidth}
\caption{Calculations of molecular dissociation and ionisation equilibria.
The predicted evolution of the partial pressure for each species --
relative to a fixed gas pressure $P_{\rm gas} = 10^3$ dyne cm$^{-2}$ --
is shown as a function of the temperature.
$P_{\rm e}$ denotes the partial pressure due to free electrons.
Two values of the C/O ratio are considered, i.e. 
the solar oxygen-rich configuration (left panel),
and a carbon-rich configuration (right panel)}
\label{fig_moltemp}
\end{minipage}
\end{figure*}
The newly developed opacity routine is then employed 
in synthetic TP-AGB evolution calculations that include
the effects of mass loss and  third dredge-up (Sect.~\ref{sec_agbmod}).
The effective temperatures of the models, 
derived through envelope integrations, are analysed as a 
function of the C/O ratio as the evolution proceeds from  oxygen-rich
to carbon-rich configurations. 
The results obtained with ``chemically-variable'' 
opacities are compared with models based on ``chemically-fixed''
opacities, as well as with observations of M- and C-type
giants in the solar neighbourhood (Sect.~\ref{sec_agbmod}).

Other possible evolutionary
effects produced by the improved opacities are discussed  
(Sect.~\ref{sec_eff}),
e.g. with respect to the C-star phase lifetimes, 
mass-loss rates, wind expansion velocities, surface carbon abundances
and yields. We also mention further related  improvements 
in the predictive capability of the models, 
i.e. to reproduce and discriminate 
the near-infrared colours of M- and C-type stars. 

Concluding comments are summarised in Sect.~\ref{sec_conc}.

\section{Opacity calculations}
\label{sec_opac}
In order to get the gas opacities for whatever distribution
of molecular species, we proceed as follows.
The mass absorption coefficient $\kappa$ (cm$^2$ gr$^{-1}$) is evaluated 
by adopting Keeley's (1970) analytical relations, 
which are parameterised as a function 
of gas density, temperature, and elemental abundances. 
These fitting formulas account for both atomic and molecular contributions 
to the total opacity. The molecules included by Keeley (1970) are:   
H$_{2}$, H$_{2}$O, OH, CO. 
The original H$_{2}$O term is modified to
reproduce more recent opacity data (see Sect.~\ref{ssec_comp}). 
We add also the 
contribution of the CN opacity by adopting the  
polynomial fit by Scalo \& Ulrich (1975). Finally, a rough estimate 
to C$_2$ opacity is given by simply assuming it as large as the CN term,  
following the results of detailed calculations by Querci et al. 
(1971; see their figure 11).

The functional form  
used to evaluate the total Rosseland mean opacity (RMO) is 
of the kind:
\begin{equation}
	\kappa= \kappa_{\rm c} + \sum_i X_i\, \kappa_i
\label{eq_k}
\end{equation}
where $\kappa_{\rm c}$ is the continuum opacity 
(due to absorption and scattering), 
$\kappa_i$ represents the RMO
produced per unit mass of molecule $i$, having a fractional abundance (by mass)
$X_i$.

Because the RMO is a harmonic (not arithmetic) mean,  in general, 
the sum of single RMO contributions 
-- as given by Eq.~(\ref{eq_k}) -- 
is not equal to the RM of the sum of monochromatic opacities. 
 Nevertheless, these two     determinations tend to converge to the same
result in case one of the involved terms 
dominates the summation.  
Actually, this condition is usually met in AGB envelopes 
under most conditions, the main source for the total RMO  
being either the continuous, or H$_{2}$O, or CN contribution
(see Scalo \& Ulrich 1975 for a discussion; see also 
Alexander et al. 1983; Alexander \& Ferguson 1994).

\subsection{Molecular concentrations}
\label{ssec_mol}

The evaluation of $\kappa$ with 
Eq.~(\ref{eq_k}) requires the abundances $X_{i}$ of the atoms and 
molecules under consideration are known. 
For given density, temperature, and chemical composition of the gas,
atomic and molecular concentrations are derived by means of 
 chemical equilibrium calculations, i.e. by solving the set
of equations describing both ionisation (Saha) and molecular 
dissociation equilibria
(see e.g. Tsuji (1966) for a description of the overall procedure).

\begin{figure}[t]
\resizebox{\hsize}{!}{\includegraphics{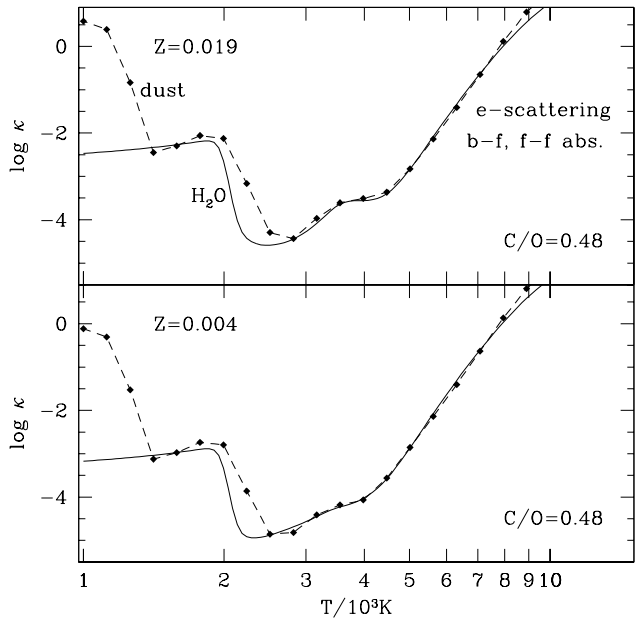}}
\caption{Mass absorption coefficient (RMO) as a function of temperature,
for two values
of the initial metallicity as labelled. For each value of $T$, the 
corresponding density is derived from the  condition 
$\log R= \log\rho + 3\, (6 - \log T) = -3$.
The C/O ratio is assumed solar (0.48) in both cases. 
The results of the present work (solid lines) are compared  
to  Alexander \& Ferguson (1994) opacity 
calculations (dashed lines; the solid squares mark the actual 
points of available data tables,  
$R-T$ interpolation being adopted in between)}
\label{fig_compz}
\end{figure}
\begin{figure}[]
\resizebox{\hsize}{!}{\includegraphics{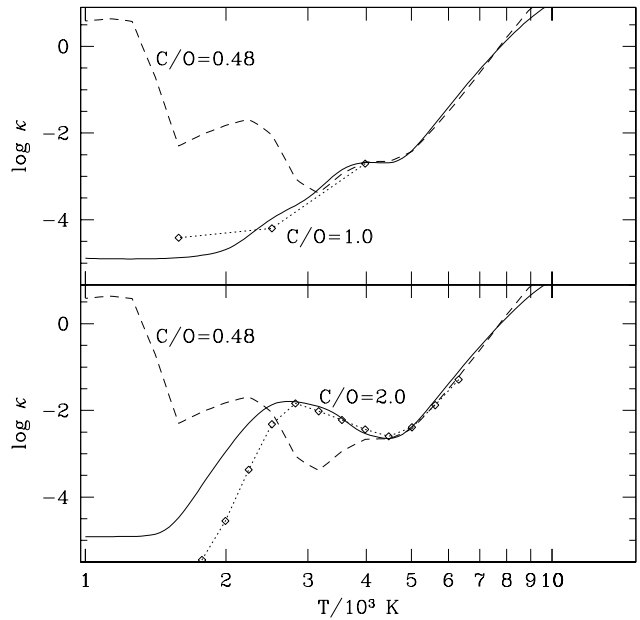}}
\caption{Mass absorption coefficient (RMO) as a function of temperature,
for different values of the C/O ratio as indicated.
In all cases the density is assumed $\log \rho=-9$. 
The results of the present work (solid lines) are compared  to
detailed opacity calculations (dotted lines) by
Alexander et al. (1983) for C/O$=1.0$ (top panel),
and Lucy et al. (1986) for C/O=$2.0$ (bottom panel).
The open squares along the curves mark the actual 
points of available data tables.
For the sake of comparison, the opacity profile according
to Alexander \& Ferguson (1994) for C/O$=0.48$ is also plotted 
(dashed line).}
\label{fig_compco}
\end{figure}

The electron pressure $P_{\rm e}$ is calculated by considering the first
ionisation stages of H, He, C, N, O, and most metals with nuclear
charge $Z$ going from 9 to 57 (the complete list is the same as in 
Alexander \& Ferguson 1994).
The partition functions for atoms are taken from Irwin (1981),
and Sauval \& Tatum (1984).
Dissociation equilibrium constants are evaluated with the aid
of the analytical expressions by Rossi \& Maciel (1983).

Figures \ref{fig_molcovar} and \ref{fig_moltemp} exemplify the results 
of standard calculations of chemical equilibrium, as a function of both 
C/O ratio and temperature.  
It is worth recalling the following
points. 

The stability of CO is so strong that,
whatever the value of the C/O ratio,  the majority of the atoms of the
least abundant element is locked up in the CO molecule, whereas the excess of
the other one can take part to the formation of other molecular
species.
This feature, pointed out long ago by Russell (1934), is responsible for 
the abrupt change in the atomic and molecular equilibria 
(involving C and O atoms) as soon as the gas mixture passes from
oxygen-rich to carbon-rich (and viceversa).

A clear example is displayed in Fig.~\ref{fig_molcovar}, where 
a remarkable increase of the fractional abundances 
(proportional to the partial pressures)
of CN and C$_2$, and a consequent drop of H$_2$O, occur as soon as  
C/O become larger than unity.
The same effect is evident by comparing the two panels of 
Fig.~\ref{fig_moltemp}, that show the expected chemical 
equilibria as a function of temperature in a mixture with the same gas 
pressure, but two 
different values of the C/O ratio. 
We also notice that for temperatures
lower than about 2000 K, the abundance of free H 
drops, as most H atoms are trapped into the H$_2$ molecule.

\begin{figure*}[t]
\begin{minipage}{0.72\textwidth}
\resizebox{\hsize}{!}{\includegraphics{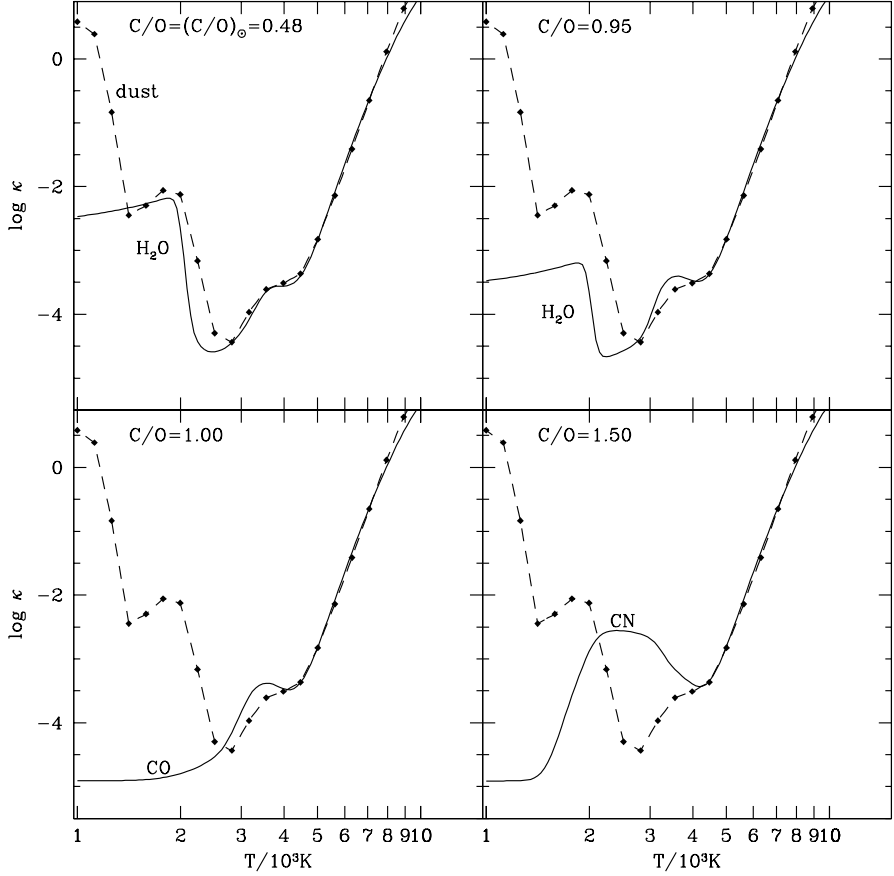}}
\end{minipage}
\hfill
\begin{minipage}{0.25\textwidth}
\caption{The same as in Fig.~\protect{\ref{fig_compz}}, but for four
different values of the C/O ratio, as indicated in each panel.
The initial metallicity is $Z=0.019$.
Results of present calculations (solid lines) are always compared
to Alexander \& Ferguson (1994; dashed lines) predictions 
that refer to fixed
solar C/O ratio. Dominant molecular contributions to the opacity
are labelled nearby the corresponding parts of the curve.   
See text for further details}
\label{fig_kcovar}
\end{minipage}
\end{figure*}

\subsection{Comparison with other opacity data}
\label{ssec_comp}
The results of the present opacity calculations are tested by 
comparison with  available data in the literature.
In Fig.~\ref{fig_compz} the comparison is performed with
Alexander \& Ferguson (1994),
whose opacity tables are largely adopted in current evolutionary
calculations of the AGB phase (e.g. Chieffi et al. 2001;  Herwig 2000; 
Ventura et al. 1999; Forestini \& Charbonnel 1997). 
In these opacity tables the abundances of metals are scaled
to the solar ones, then implying a C/O ratio of about 0.48 for any
value of the metallicity.
Figure~\ref{fig_compz} illustrates two examples, that refer to 
two choices of the  initial composition, namely: [$Y=0.273$, $Z=0.019$] and
[$Y=0.240$, $Z=0.004$].
In both cases, for each temperature value, 
the associated density satisfies the
condition $\log R= \log\rho + 3\, (6 - \log T) = -3$, where 
$R$ is a commonly adopted quantity as it  
allows smooth interpolations of opacity tables.

Our predictions in Fig.~\ref{fig_compz} show a general agreement 
with Alexander  \& Ferguson (1994), reproducing rather well the main features
of the opacity for temperatures in the range from about 
$10\,000$ K down to about $1\,500$ K.
In particular, we account for 
the first opacity bump at $T \la 2\,500$ K due to water vapor.
To this respect, we should also note that, according to Alexander  
\& Ferguson (1994), an additional smaller contribution to the opacity 
bump comes from TiO, which is not included in the present treatment.    
 
Anyhow, as the original (Keeley 1970) formula  
underestimates the H$_2$O opacity compared to Alexander  \& Ferguson's 
results, we choose to modify the corresponding
 analytical term $k_{\rm H_{2}O}$
in Eq.~(\ref{eq_k})\footnote{More specifically, the numerator 
$9.72 \times 10^{-18} e^{-3.2553/T_4}$  has been replaced 
with $9.72 \times 10^{-21} 
e^{-3.2553/(T_4+0.37)}$, where $T_4 = T/10^4\, K$}, 
so as to obtain 
a better reproduction of the more recent opacity data. 
In fact, the water vapor opacity has been
totally revised over the years, essentially following the 
introduction of new techniques and extension of the adopted line list
(see e.g. Alexander et al. 1989; Gustafsson 1995).

The second opacity bump for $T < 1\,500$ is not reproduced by our 
calculations as the dust contribution is not taken into account.
However, these very low temperature are not attained in the 
atmospheres of our carbon star models.  
 
Figure~\ref{fig_compco} compares our results with available opacity data 
for gas mixtures with C/O$=1$ (Alexander et al. 1983), and C/O$=2$ 
(Lucy et al. 1986). 
The oxygen abundance is kept solar, while
that of carbon is increased so as to obtain the specified C/O ratio.
We notice in both cases the opacity curves at 
$T < 4\,500$ K largely deviate from that corresponding 
to a solar-scaled mixture of metals (with C/O$\sim 0.48$) 
(see the discussion below in Sect.~\ref{ssec_kco}).
The basic features -- i.e. the disappearance  of the H$_2$O opacity 
bump for C/O$=1$ and the appearance of the CN opacity bump 
for  C/O$=2$ -- present in the data tables are also predicted  
by our calculations. Some differences exist, like a possible
overestimation of the CN opacity at temperatures lower than 
$2\,500$ K (bottom panel of Fig.~\ref{fig_compco}).
Nevertheless, the results shown 
in Figs.~\ref{fig_compz} and \ref{fig_compco}
clearly show that 
our simple opacity treatment provides a satisfactory description
of the behaviour of molecular opacities with varying C/O ratio, 
and represents a better alternative to the usual  
assumption of opacities for solar-scaled metal abundances.

\subsection{Molecular opacities for variable C/O ratio}
\label{ssec_kco} 
Figure~\ref{fig_kcovar} shows a striking example of how large might 
be the discrepancy  between  ``chemically-fixed'' 
and ``chemically-variable'' opacity data. 
The former case refers to a fixed solar-like composition 
according to Alexander \& Ferguson (1994), whereas the latter case
corresponds to a chemical mixture with increasing C abundance.

We can see that major variations of the opacity
show up as the C/O ratio increases 
from below to above unity (due to the progressive
increase of the C abundance). 
At lower temperatures ($T \la 3\,000$ K)
the H$_2$O bump drops until it disappears at C/O$\sim 1$, as
almost all atoms of both C and O are bound in the
CO molecule.
Then, as soon as C/O overcomes unity,
the CN$+$C$_2$ opacity contribution suddenly rises, becoming
the dominant opacity source at temperatures between $2\,000$ and $3\,000$ K.
It turns out that such a prominent opacity bump develops just where
an opacity minimum is instead expected for oxygen-rich mixtures.

Therefore, by comparing the results displayed in Fig.~\ref{fig_kcovar}, 
it is already clear that applying the opacity profile 
expected for a solar mixture to calculate the atmosphere
of a carbon-rich model is a considerable mismatch.
However, this is presently the usual choice in stellar evolution calculations
of the AGB phase.  

\subsection{Pros and cons of the present opacity treatment}

A few comments should be made with respect to the pros and cons
of the present treatment. A clear limitation is related to
the adopted simplifications, namely: calculate $\kappa$ as the sum
of RMOs of individual molecules, and include    
a limited number of molecular species.
However, we already mentioned that Eq.~(\ref{eq_k})
may be considered an acceptable approximation under most conditions
met in AGB atmospheres. 
Possibly important molecules in AGB stars, that are neglected in this
study, are TiO and VO for oxygen-rich stars and HCN and C$_2$H$_2$ 
for carbon-rich stars. Anyhow, the molecules here considered  
are among the most relevant opacity sources at low temperatures in AGB stars.

The major advantage is offered by the possibility to evaluate
the opacities for {\sl any} chemical composition, {\sl during}
the evolutionary calculations, and with just a small additional computational
effort. 

Clearly, a more accurate alternative 
would be  to perform multidimensional interpolations between tables 
previously generated by detailed opacity calculations.
However, it should be remarked that in order to guarantee a
sufficient coverage of the possible conditions met in AGB atmospheres,
a large grid of tables is needed
(and presently not available), for many  combinations of 
various parameters. In the specific case of AGB stars, 
we should deal with a minimum of seven parameters, namely:
density, temperature, total metal content, hydrogen, carbon, nitrogen, and
oxygen abundances.

It follows that, despite the involved approximations, 
our approach is a reasonable compromise and also  
an improvement upon commonly adopted input prescriptions 
in AGB evolution models (i.e. solar-scaled molecular opacities),
as it can describe the changes in the atmospheric opacities 
consequent to the changes in the surface chemical composition of AGB
stars.

 
\section{Synthetic TP-AGB calculations}
\label{sec_agbmod}
The new routine to evaluate the molecular opacities, described 
in Sect.~\ref{sec_opac}, is incorporated in the envelope model employed
in the code for synthetic TP-AGB evolution
developed by Marigo et al. (1996, 1998) and  
Marigo (2001, and references therein),
to whom the reader is referred for all details.
 
Let us here summarise just the basic structure of the 
TP-AGB synthetic model. It consists of two main components, namely:
\begin{itemize}  
\item  analytical ingredients  derived from full stellar calculations 
and/or observations (e.g.
core mass-luminosity relation for low-mass stars, core mass-interpulse 
period relation, mass-loss and dredge-up laws, etc.); 
\item a complete ``extended'' envelope model that, at each time step, 
 numerically solves the structure of the stellar atmosphere and
underlying convective envelope, down to the H-burning shell.
\end{itemize}

As for the envelope model, the solution scheme is the following.
Given total stellar mass, core mass, 
and surface chemical composition, 
the stellar structure equations are integrated to  
determine the profile of the unknown variables 
$r$, $P_{r}$, $T_{r}$, $L_{r}$ across the envelope, 
provided that two outer (at the photosphere) and 
two inner (at the core) boundary conditions are fulfilled.
In particular, the outer boundary conditions --  derived from 
the integration of the photospheric equations 
for $T$ and $P$ --  are related to the  atmospheric parameters
$L$ and $T_{\rm eff}$.
Static and grey atmospheres are assumed. 
Convection is described according  to 
the mixing-length (ML) theory, and the ML parameter is 
$\alpha=\Lambda/H_{P} = 1.68$ (where $\Lambda$ is the mixing length, and 
$H_{P}$ is the pressure-scale height).

Envelope integrations are carried out 
to predict the evolution of the envelope structure and
surface properties of a TP-AGB star during the
quiescent inter-pulse periods, that is when the H-burning
shell provides most of the stellar energy, while the  
He-shell contribution is small. 
In terms of duration, this quiescent phase -- 
between the occurrence of two consecutive thermal pulses -- 
is by far dominant, given the extremely shorter time over which  
a thermal pulse takes place.

The envelope model is also employed to test the possible occurrence of the
third dredge-up, according to a criterium on the temperature at the base 
of the convective envelope in the stage of post-flash luminosity maximum. 
The minimum base temperature required for dredge-up to
take place is $\log T_{\rm b}^{\rm dred} = 6.4 $, following the 
empirical calibration performed by Marigo et al. (1999; see also Wood 1981) 
on the basis of the observed carbon star luminosity functions 
in the Magellanic Clouds.
The efficiency of the third dredge-up, 
expressed by the classical quantity 
$\lambda$, is a free parameter, that varies within 0.50-0.75 
in the present calculations.
The recurrent dredge-up of carbon at thermal pulses is responsible
for the transition of the models  from the (C/O$< 1$) to the (C/O$>1$) domain.
Hot-bottom burning (and the related break-down of the core mass-luminosity
relation) in the most massive models  (with $M \ga 3-4\, M_{\odot}$) 
is also taken into account (Marigo et al. 1998, Marigo 1998).  
The TP-AGB evolution is calculated  with the inclusion of mass loss
up to the complete ejection of the residual envelope. 
The adopted semi-empirical formalism for mass-loss is that developed by 
Vassiliadis \& Wood (1993).

For the purposes of this study, a limited set of 
synthetic TP-AGB models has been calculated,  
with initial metallicity $Z=0.019$ and masses 
in the range $1.2\, M_{\odot}\, - \, 3\, M_{\odot}$.
For each stellar model, calculations start at the first thermal pulse,
following the predictions of the Padova stellar 
tracks (Girardi et al. 2000), whence we extract our
initial conditions (i.e. core mass, envelope chemical composition,
luminosity, etc.). 
In the TP-AGB models here considered, with relatively low masses,
 the surface chemical composition may be altered 
by the third dredge-up only, whereas hot-bottom burning does not
take place.  

As far as the adopted opacity prescriptions are concerned, we
distinguish two groups of models.
In both cases we take the tables  by   
Iglesias \& Rogers (1993) for $\log T > 4.0$, and 
Alexander \& Ferguson (1994) for $3.7 < \log T \le 4.0$.
The only  difference resides in the choice 
of the adopted opacities for $\log T \le 3.7$: 
\begin{itemize}
\item models $F$: fixed opacities
for solar-scaled metal abundances
\item models $V$: variable opacities (Sect.~\ref{sec_opac}) coupled to the 
changes in chemical abundances (e.g. C and O) due to dredge-up events
\end{itemize}
In fact, these are the typical temperatures at which the 
molecular concentrations become appreciable
(see e.g. Figs.~\ref{fig_compz} -- \ref{fig_kcovar}).

\begin{figure*}
\begin{minipage}{0.72\textwidth}
\resizebox{\hsize}{!}{\includegraphics{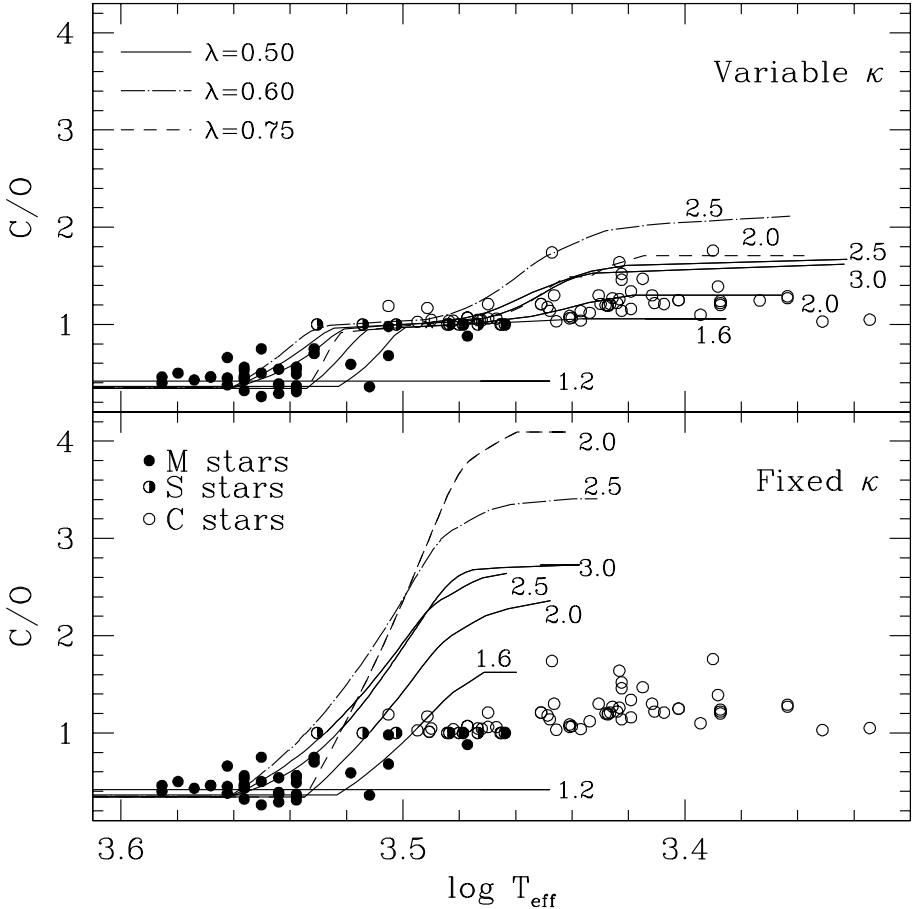}}
\end{minipage}
\hfill
\begin{minipage}{0.25\textwidth}
\caption{Effective temperatures as a function of the C/O ratio
in  Galactic giants.
Abundance determinations are taken from:
Smith \& Lambert (1985, 1986, 1990) for M stars  (C/O$ <  1$);
Ohnaka \& Tsuji (1996) for S stars (C/O$\sim 1$);
Lambert et al. (1986), Ohnaka et al. (2000) for C stars (C/O$ >  1$).
Effective temperatures are taken from the same quoted works,
except the C-star group for which we refer to  
Bergeat et al. (2001; their table 13).
Observed data (circles) are compared to predictions of synthetic
TP-AGB models with dredge-up (lines), adopting either ``chemically-fixed'' 
opacities (bottom panel), or ``chemically-variable'' opacities (top panel).
See text for more details}
\label{fig_cotef}
\end{minipage}
\end{figure*}

\section{Effective temperatures of oxygen- and carbon-rich stars}
\label{sec_cotef}
It has long been recognised that the opacity of the atmospheric
layers plays a crucial role in determining the surface properties of
giant stars, in particular affecting their position on the H-R diagram
(e.g. Lucy et al. 1986).
From stellar models we learn that 
for given luminosity (determined e.g. by the core mass-luminosity
relation), larger atmospheric  opacities 
usually correspond to larger radii, hence lower effective
temperatures.

This may apply for instance, to a C-rich atmosphere 
compared to an O-rich atmosphere, if the typical range 
of involved temperatures across both atmospheres is 
(say between $2\,000$ and $4\,000$ K) is such to include important 
opacity features, like the CN opacity bump (or opacity minimum) 
of the C-rich (or O-rich) configuration (see Fig.~\ref{fig_compz}).  
In this case, then, a cooler effective temperature should 
describe the C-rich model.

This prediction may explain, in fact, the existence of a clear relation
between the effective temperatures of giant stars and their surface C/O ratios:
Carbon-rich stars are found to have lower effective temperature
than oxygen-rich stars. 
This trend is  illustrated in Fig.~\ref{fig_cotef},
where the observed data refer to a sample of Galactic giants.

The effective temperatures 
of most carbon stars of the sample are the most accurate possible 
determinations, as they have been derived with the aid of
{\sl direct} methods, based on angular diameters estimated with
interferometry or lunar occultations (see the compilation 
by Bergeat et al. (2001) and references
therein). For the other stars (of M-type with C/O$<1$ and SC-type with
C/O$\sim 1$), 
effective temperatures are indirectly derived via spectral
analysis techniques employed to determine their CNO abundances, 
hence C/O ratio (e.g. Smith \& Lambert 1986).

\begin{figure*}
\begin{minipage}{0.72\textwidth}
\resizebox{\hsize}{!}{\includegraphics{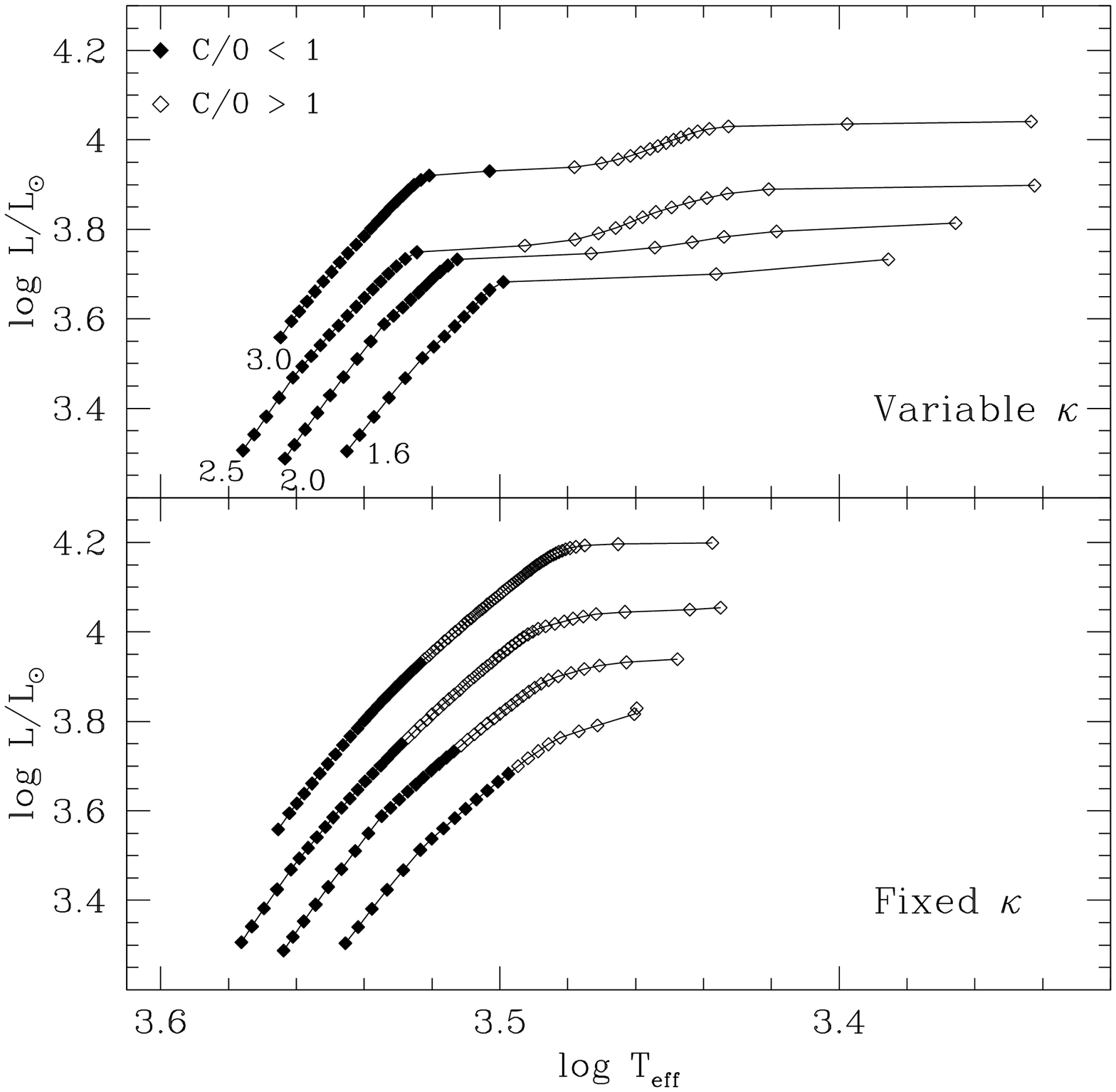}}
\end{minipage}
\hfill
\begin{minipage}{0.25\textwidth}
\caption{Predicted H-R tracks of TP-AGB models 
with initial metallicity $Z=0.019$.
Calculations are carried out with both
fixed (bottom panel) molecular opacities for solar composition, 
and variable opacities (top panel)
related to the current photospheric abundances of C and O 
(hence C/O ratio) during the evolution. In all models the dredge-up
efficiency is assumed $\lambda=0.5$. Squares along the tracks mark
the pre-flash luminosity maximum at the end of each pulse-cycle}
\label{fig_hr}
\end{minipage}
\end{figure*}

The empirical relation shown in Fig.~\ref{fig_cotef} clearly indicates 
two major facts, namely: i) the almost 
complete segregation  in effective temperature between 
oxygen-rich and carbon-rich stars; and ii) the relatively low C/O values
($< 2$) measured in carbon-rich stars.

The observed data are then compared to the two sets 
($F$ and $V$) of TP-AGB synthetic models, that only differ
in the adopted prescriptions for low-temperature opacities 
($\log T < 3.7$).

Figure~\ref{fig_cotef} shows the remarkable  
disagreement between observations and predictions for models 
$F$ (bottom panel). 
The observed domain of C-stars is not reached by the 
stellar tracks, that are characterised by higher effective temperatures
and larger C/O ratios.
It should be also noticed that, as the surface C/O ratio increases 
because carbon is added
at each dredge-up episode, the decrease of
$\log T_{\rm eff}$ follows an almost straight line, up to when a more
significant bending towards lower $T_{\rm eff}$ takes place during
the very last quiescent inter-pulse periods.
Actually, the slope of the C/O$ - \log T_{\rm eff}$ tracks 
does not change even when models $F$ make the transition from C/O$< 1$ 
to C/O$>1$. This 
can be understood just from the fact that in models $F$ the molecular 
opacities are not affected at all by changes in the CNO abundances.
The final flattening of  tracks is instead caused by the drastic 
reduction of the envelope mass at the onset of the super-wind regime. 
 
The effect of the new molecular opacities shows up sharply 
in models $V$.
First of all, they succeed in  reproducing the observed location 
of both oxygen- and carbon-rich stars, hence removing 
the aforementioned discrepancy.
The main aspect is that, contrary to models $F$,  
models $V$ with variable opacities perform quite a large excursion
towards lower effective temperature as soon as their surface abundance 
of C exceeds that of O.
Such excursion is initially driven by the sudden appearance
and progressive build-up of the molecular opacities 
due to CN and C$_2$ 
(see Figure~\ref{fig_kcovar}) as the C/O ratio becomes larger than unity.
The photospheric cooling, in turn, favours larger and larger mass-loss rates,
contributing to anticipate the super-wind phase.

In brief, the behaviour of tracks $F$ and $V$ notably differs in the 
flattening towards lower $T_{\rm eff}$ shown in Fig.~\ref{fig_cotef}.  
In models $F$ the bending is caused by the reduction of the envelope mass, and 
starts at the onset of the  superwind {\sl after} the transition to the 
C-rich class.
In models $V$ it is initiated by the drastic change in molecular opacities
{\sl as soon as} the oxygen- to carbon-rich transition occurs, and
subsequently amplified during the super-wind phase.
This feature can be seen also by looking at the TP-AGB evolution in the 
H-R diagram (see Fig.~\ref{fig_hr}): 
compared to models $F$ with the same initial mass,
the flattening of tracks $V$ occurs at fainter luminosities and
extends to cooler $T_{\rm eff}$.

It is also interesting to notice that the transition to the C-star
configuration is characterised by a first sizeable jump towards
lower $T_{\rm eff}$ (compare the last C/O$<1$ point with the first
C/O$>1$ point along each H-R track of group $V$), 
which is more pronounced at decreasing 
stellar mass. After this initial sharp departure away 
from the oxygen-rich part 
of the H-R track, the subsequent cooling  
proceeds at a slower rate during the C-star evolution, as illustrated
by the denser sequence of points marked along tracks $V$ 
in Fig.~\ref{fig_hr}. Finally, a further acceleration towards
lower $T_{\rm eff}$ takes place as soon as the super-wind is attained
and the envelope is rapidly ejected 
(correspondingly, the distance in $T_{\rm eff}$ 
between the last points increases).

This predicted behaviour naturally succeeds in explaining the observed 
spread and distribution in effective temperature exhibited by 
the sample of M-S-C stars shown in Fig.~\ref{fig_cotef}. 
To this aim, let us consider the
overall morphologic evolution of the bundle of theoretical TP-AGB tracks
(models $V$, top panel).

For C/O$<1$ (corresponding to M-stars),
the width in $T_{\rm eff}$ of the bundle  
is relatively narrow and reflects  
individual variations of fundamental parameters  
(essentially: envelope and core masses, and 
dredge-up law). 
  
For C/O$\sim 1$, at the M-to-C transition, the bundle   
stretches out over 
a rather extended $T_{\rm eff}$-interval, 
as a consequence of the expected jump shown by each track. 
This is fully supported
by the observed location of S-stars in Fig.~\ref{fig_cotef}. 

Finally, as the bundle widens for C/O$>1$, 
the evolution is characterised by a further significant cooling of the tracks,
that depends on individual stellar properties    
(compare e.g. the tracks of the same stellar mass but with 
different $\lambda$). 
Indeed, models $V$ are in agreement with the empirical indication, 
already pointed out by Bergeat et al. (2001), that 
the dispersion of the C/O values measured in C-stars increases 
at decreasing effective temperatures.

\section{Other evolutionary and observational effects}
\label{sec_eff}
\begin{table*}
\begin{minipage}{0.89\textwidth}
\caption{Predictions of synthetic TP-AGB models calculated with 
fixed (F) molecular opacities for solar composition, 
and variable (V) molecular opacities. The assumed dredge-up
efficiency parameter is $\lambda=0.5$.
From left to right the table entries correspond to: stellar initial
mass; duration of the TP-AGB phase; duration of the C-star phase; 
 mean effective temperature when the surface C/O$\sim 1.1$; and
values of C/O, luminosity, and final mass 
at the termination of the AGB}
\label{tab_co}
\begin{tabular}{ccccccccccccc}
\noalign{\smallskip}
\hline
\noalign{\smallskip}
\multicolumn{1}{c}{$M_{\rm i}/M_{\odot}$} &
\multicolumn{2}{c}{$\tau_{\rm TP-AGB}/(10^6$ yr)} &
\multicolumn{2}{c}{$\tau_{\rm C}/(10^6$ yr)} &
\multicolumn{2}{c}{$\langle T_{\rm eff}\rangle({\rm C/O}\sim 1.1)$} &
\multicolumn{2}{c}{(C/O)$_{\rm f}$} &
\multicolumn{2}{c}{$\log L_{\rm f}/L_{\odot}$} & 
\multicolumn{2}{c}{$M_{\rm f}/M_{\odot}$} \\ 
\noalign{\smallskip}
\hline
\noalign{\smallskip}
\multicolumn{1}{c}{} &
\multicolumn{1}{c}{F} &
\multicolumn{1}{c}{V} &
\multicolumn{1}{c}{F} &
\multicolumn{1}{c}{V} &
\multicolumn{1}{c}{F} &
\multicolumn{1}{c}{V} &
\multicolumn{1}{c}{F} &
\multicolumn{1}{c}{V} &
\multicolumn{1}{c}{F} &
\multicolumn{1}{c}{V} &
\multicolumn{1}{c}{F} &
\multicolumn{1}{c}{V} \\
\noalign{\smallskip}
\hline
\noalign{\smallskip}
1.6 & 1.94 & 1.46 & 0.71 & 0.23 & 3\,102  &  --- &  1.62 & 1.06 
& 3.829 & 3.746 & 0.596 & 0.577 \\
2.0 & 3.37 & 2.41 & 1.48 & 0.51 & 3\,228 & 2\,830 & 2.36 & 1.30 
& 3.939 & 3.814 & 0.632 & 0.588 \\
2.5 & 3.67 & 2.69 & 1.92 & 0.94 & 3\,341 & 2\,972 & 2.64 & 1.67 
& 4.054 & 3.907 & 0.686 & 0.621 \\
3.0 & 2.91 & 1.92 & 1.67 & 0.68 & 3\,311 & 2\,929 & 2.73 & 1.62 
& 4.209 & 4.048 & 0.780 & 0.685 \\
\noalign{\smallskip}
\hline
\end{tabular}
\end{minipage}
\end{table*}

In addition to the mentioned effects on the effective temperature, 
the use of opacity prescriptions 
properly coupled to the current CNO abundances, may bring along  
other  evolutionary implications of great relevance.

\begin{figure}
\resizebox{\hsize}{!}{\includegraphics{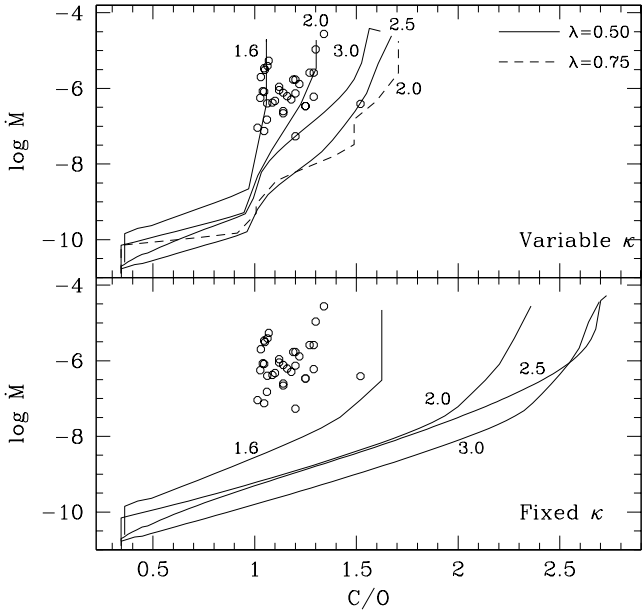}}
\caption{Mass-loss rates on the AGB as a function of the surface C/O
ratio. Observed data for carbon stars (circles) combine 
$\dot M$ determinations by Loup et al. (1993), 
and Wannier et al. (1990) with C/O ratio derived by Lambert et al. (1986), 
and Ohnaka et al. (2000). Theoretical tracks (lines) show the evolution
of $\dot M$ as the surface C/O increases because of the third dredge-up.
By comparing the top and bottom panels we can note 
how different are the predictions for the carbon-enriched models
(with C/O$>1$), depending on the molecular opacities adopted in 
envelope integrations }
\label{fig_mdot}
\end{figure}

\begin{figure}
\resizebox{\hsize}{!}{\includegraphics{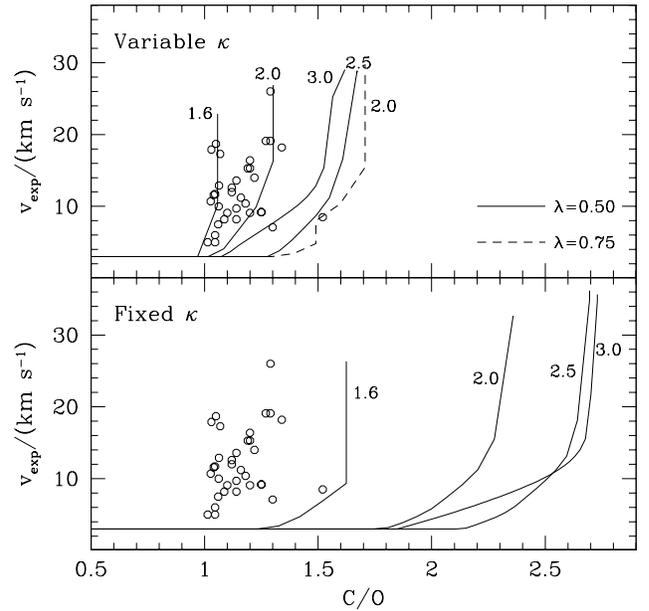}}
\caption{Expansion velocities of AGB ejecta as a function of the surface C/O
ratio. Observed data for carbon stars (circles) combine
$v_{\rm exp}$ determinations by Loup et al. (1993) with C/O ratio 
derived by Lambert et al. (1986), Ohnaka et al. (2000).
Theoretical tracks (lines) refer to the same sets of models as 
in Fig.~\protect\ref{fig_mdot} }
\label{fig_vexp}
\end{figure}

Some indications are presented in Table \ref{tab_co} 
and Figs.~\ref{fig_mdot}, \ref{fig_vexp} that compare
the predictions for some representative quantities 
of synthetic TP-AGB models
calculated with the same input physics but for the molecular
opacities (i.e. models $F$ and $V$ presented in Sect.~\ref{sec_cotef}).

First of all, we can notice in Table \ref{tab_co} that the introduction
of the new opacities in our TP-AGB calculations 
has determined  a 
sizeable reduction of the duration of the C-star phase, 
by a factor 2-3 for the specific models under consideration. 

As already mentioned, such effect is related to the excursion 
towards lower effective 
temperatures caused by the transition from C/O$<1$ to C/O$>1$.
This, in turn, favours larger mass-loss rates, hence anticipating 
the onset of the super-wind regime and the consequent 
termination of the AGB phase.

Another consequent effect concerns the luminosities 
of C-stars and the final masses at the termination of the AGB.  
As shown in Fig.~\ref{fig_hr}, the luminosity interval spanned 
by models $V$ for C/O$>1$ is much smaller than in models $F$ with
the same mass and assumed dredge-up efficiency.
Correspondingly, the AGB-tip luminosities, $L_{\rm f}$,  
and the final masses, $M_{\rm f}$,  
are lower in models $V$ than in models $F$ (see Table~\ref{tab_co}). 
From inspection of Fig.~\ref{fig_hr} (top panel), 
one could speculate that a well-defined
relation exists between the observed C-star luminosity and its initial 
mass, and hence the age of its parent population. Moreover, since
models $V$ leave the AGB phase earlier than models $F$,
we might expect that the initial-final mass relation tends to flatten over 
the initial-mass interval pertaining to the C-stars' progenitors.  
Before drawing any conclusion
on these issues, however, it is necessary 
to perform an empirical re-calibration of the other model parameters
(e.g. dredge-up law) that also affect the predicted C-star luminosities and
final masses. 
The basic observables
to be reproduced are the C-star luminosity functions 
e.g. in the Magellanic Clouds, and the white-dwarf mass distribution, 
as done by Marigo et al. (1999).

\begin{figure*}
\begin{minipage}{0.72\textwidth}
\resizebox{\hsize}{!}{\includegraphics{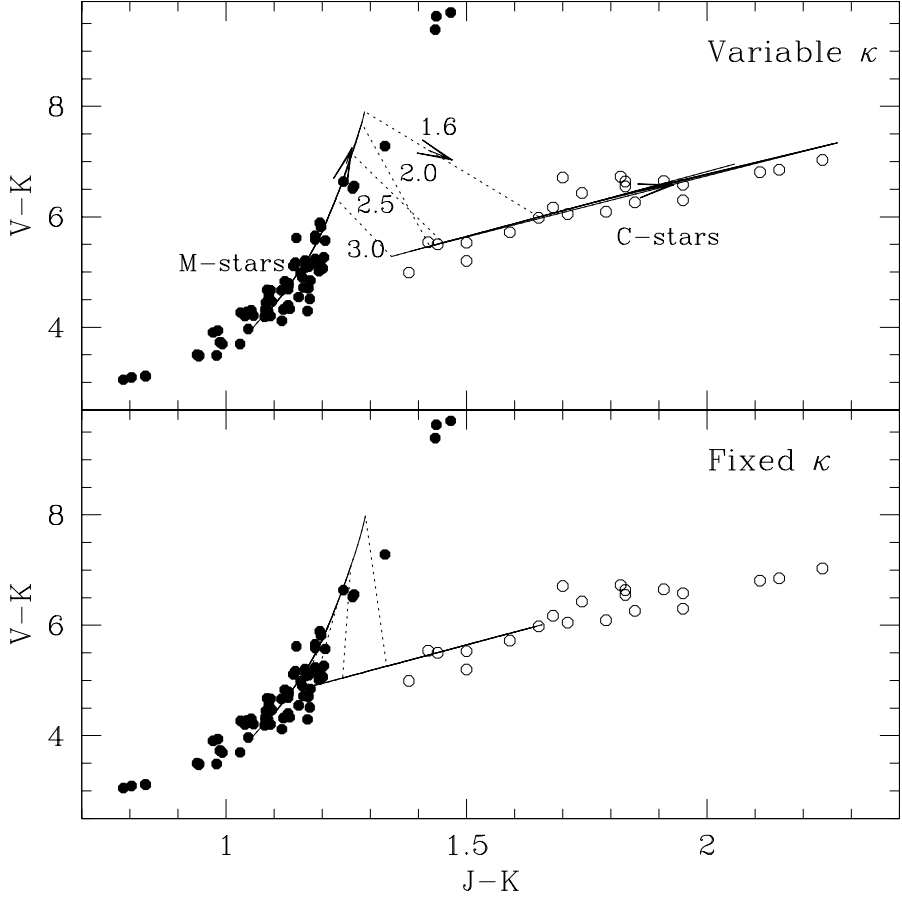}}
\end{minipage}
\hfill
\begin{minipage}{0.25\textwidth}
\caption{Near-infrared colour-colour diagram for oxygen-rich
(M-type) and carbon-rich (C-type) stars in the Solar neighbourhood.
Observed data are taken from the compilations 
of M-stars by  Fluks et al. (1994; filled circles), 
and  C-stars by Bergeat et al.
(2001, objects in their table 4; empty circles).
The predicted colour evolution on the TP-AGB is
shown (solid lines, the direction is indicated by arrows) 
for different initial stellar masses (indicated in $M_{\odot}$),
and fixed/variable molecular opacities. See text for 
more explanation
}
\label{fig_color}
\end{minipage}
\end{figure*}

With respect to mass loss on the AGB, we can see in Fig.~\ref{fig_mdot}
that models $V$ describe the behaviour of the observed mass-loss rates 
of C-stars as a function of the C/O ratio remarkably well.
Similarly to the H-R tracks, also the tracks $V$ in the 
($\dot M$ -- C/O) plane show an evident change of slope 
as soon as surface carbon
becomes more abundant than oxygen, with a clear acceleration towards 
larger and larger $\dot M$. 
Differently, in models $F$ the change of slope is found to occur
at the onset of the superwind, that is after the transition into
the C/O$>1$ domain.
An analogous situation applies to the evolution of the terminal (or expansion) 
velocity, $v_{\rm exp}$, of the AGB wind as a function of the C/O ratio, 
which is shown in Fig.~\ref{fig_vexp}.
It is calculated -- following Vassiliadis \& Wood's (1993) prescriptions --
as a function of the pulsation period which, in turn, depends also
on $T_{\rm eff}$.
Again, models $V$ are in much better 
agreement with observations of C-stars compared to models $F$, and again
a sudden change of slope is seen at the C-star formation.
All these features point to the same conclusion: the surface C/O
ratio of AGB stars is a crucial
factor in determining their evolutionary properties.

Further notable consequences arise from this ascertainment. 
For instance, the reduction of the C-star lifetimes 
(passing from models $F$ to models $V$) consequently 
affects other model predictions regarding e.g.
the expected $N$(C)/$N$(M) ratio between the number of
C- and M-stars 
in old- and intermediate-age  stellar populations, the carbon abundances 
and yields. Basing on the results presented in 
Table~\ref{tab_co}, all these quantities show a decreasing trend 
going from models  $F$ to models $V$. 
For instance, the ratio $\tau_{\rm C}/(\tau_{\rm TP-AGB}-\tau_{\rm C})$
as a function of stellar mass is a measure of the expected 
$N$(C)/$N$(M) ratio -- limited to late M-stars -- 
as a function of the age of the corresponding 
simple stellar population.
According to present calculations, this ratio 
varies in the interval
0.58 -- 1.35 for models $F$, and 0.19 -- 0.55 for models $V$.
Moreover, the interesting possibility of a lower production (and ejection) 
of carbon might help to explain the rather low
C/H and C/O abundances measured in PNe, such as the accurate 
determinations recently derived from ISO spectra (Pottasch 2000).
Actually, lower stellar yields of carbon seem also required by 
chemical evolution models of galaxies 
to better reproduce the observations (Portinari et al. 1998).

%
The improved treatment of molecular opacities should also allow a better 
description of near-infrared colours -- mainly in the $JHK$ bands 
-- of AGB stars.
Leaving a more detailed analysis of this topic to a future
investigation, we briefly mention the observed 
dichotomy between oxygen- and carbon-rich stars in the
($J-K$) vs. ($V-K$) diagram.
As shown in Fig.~\ref{fig_color}, M- and C-stars populate two different
branches, the carbon-rich objects exhibiting  systematically redder
($J-K$) colours.   

The predicted evolution of the colours is also 
shown for both sets of models. The transformations from 
the theoretical to the observational plane are performed by using
the calibrated colour-temperature relations presented by
Fluks et al. (1994) for the M-stars, and Bergeat et al. (2001) for
the C-stars.
Combining the two transformations, an abrupt jump in both 
($J-K$) and ($V-K$) colours
is expected as soon as the stars pass from 
oxygen-rich to carbon-rich (dotted lines).

As for the C-star branch, we notice that models
with variable opacities well extend throughout the observed range 
of the ($J-K$), from about 1.3 up to  about 2.2.
Differently, carbon-rich models with fixed opacities 
get mixed into the M-star domain at lower colours 
(contrary to observations), 
and draw a shorter evolution
towards redder colours, not exceeding ($J-K) \sim 1.7$.

It is clear that a deeper analysis of the results presented
in this work requires extensive calculations 
of TP-AGB models, coupled to a close comparison with 
observations. 
It should be also considered that any further change in the input
prescriptions -- besides those for the molecular opacities --  would
produce additional effects. It follows that, 
in order to reproduce basic observational constraints 
(like the carbon star luminosity functions and the white dwarf 
mass distribution; see e.g. Marigo 2001), 
the adoption of the improved opacities should be accompanied by     
a re-calibration of the relevant model
parameters (mainly dredge-up law and mass-loss efficiency). 
All these aspects will be considered with more detail in future work.

Anyhow, just from basing on the present explorative calculations, 
there is a clear hint that the new
variable opacities go in the right direction to match
theory and observations.

\section{Concluding remarks}
\label{sec_conc}
In this work we call attention to the importance of consistently coupling
the molecular opacities with the current envelope chemical composition 
in TP-AGB evolution models, and to discontinue the use of 
opacity tables that are generated for fixed -- both absolute and relative 
(usually solar-scaled) -- abundances of metals, 
like carbon and oxygen.

Our explorative study starts with the development of a routine to
estimate the molecular opacities for any choice of the chemical
composition of the gas.
Notwithstanding the unavoidable approximations, 
this tool is easily incorporated
in our envelope model, that is employed  
to derive the effective
temperature of TP-AGB models during their quiescent inter-pulse evolution. 
In this way, we can follow  
the evolution of molecular opacities as more carbon 
is dredged-up to the surface, and in particular the abrupt change
in the dominant opacity sources at the transition from the
the oxygen- to the carbon-rich domain.
Correspondingly, the most remarkable consequence is the sudden cooling 
of the stellar tracks in the $\log L - \log T_{\rm eff}$ diagram, which 
causes  the displacement of the C-rich models towards redder 
near-infrared colours.

We expect that the latter effect should depend, among other factors, 
on the global metal content of the stellar population which the 
observed AGB stars belong to. 
Preliminary calculations suggest that the cooling of the C-rich models 
away from the O-rich AGB should be less pronounced at decreasing
metallicity, possibly becoming even negligible in AGB populations of very
low metallicity. This would occur if the 
temperatures across the atmosphere are too warm (say $> 4\,000$ K, 
or equivalently for $\log T_{\rm eff} > 3.6$) 
and  do not fall in the temperature range 
required for the formation of the characteristic molecules 
(e.g. CO, H$_{2}$O, and CN) in appreciable concentrations.

Furthermore, the results obtained in this work indicate that  
-- while keeping fixed all other
prescriptions -- the adoption 
of more consistent molecular opacities in AGB models  
produces significant effects, such as:  
shortening of the C-star phase, lower AGB-tip luminosities and final 
masses for C-stars, lower carbon yields. 
Of course, all these aspects need to be investigated in the context 
of a more extended analysis that performs a re-calibration 
of the model parameters (e.g. the dredge-up law) 
to reproduce basic observables of C-stars (Marigo et al., in preparation).

It is also clear that the predicted evolution of the effective temperature 
as a function of the surface C/O ratio -- shown in this study 
for AGB models with initial solar metallicity -- should
be considered representative of a mean trend, as 
it refers to quiescent inter-pulse stages.
Additional effects -- such as stellar pulsation (instead of the assumed
staticity of the envelope), and the complex variation of 
the envelope structure (in terms of contraction and/or
expansion) driven by thermal pulses -- should produce some 
further dispersion around the mean $T_{\rm eff}$ at given C/O.
Moreover, major improvements in the opacity treatment 
could be adopted in future work, including the resort to detailed
opacity libraries as they may become available.  

Anyhow, we expect that none of these additional elements could 
change the main point
of this study: the large inadequacy of fixed solar-scaled molecular
opacities in AGB models with varying C and O 
surface abundances, and the marked cooling of the AGB for C/O$>1$.

Actually, a net progress in model predictions is already attained
in the present analysis, as supported by the
first successful comparisons with observations of e.g. C/O ratios,
effective temperatures, and near-infrared  colours of C-stars.
This improvement in modelling the TP-AGB phase 
opens the way to promising applications,
aimed at investigating the contribution of old- and
intermediate-age stellar populations to the chemical and 
spectro-photometric evolution of the parent galaxies.

More realistic TP-AGB models 
are really needed to keep up with observational advancements: 
AGB stars have been resolved  in
many galaxies of the Local Group 
(see e.g. Nowotny et al. 2001; Saviane et al. 2000;  
Groenewegen 1999 for a recent census), 
and also indirectly revealed
via their signature in the integrated spectra of the stellar
systems (e.g. Lan\c con et al. 1999).
Moreover, huge amounts of infrared data 
with unprecedented sky coverage are being released 
(e.g. the DENIS and 2MASS projects).
 
In this context, the issue addressed in this work 
should provide an important step
forward to bring AGB stellar models closer to observations.

\begin{acknowledgements}
I would like to thank   
Harm Habing for fruitful suggestions and remarks on this work, 
and L\'eo Girardi for careful and critical reading 
of the manuscript.
This work is financially supported by the Italian 
Ministry of Education, University and Research (MIUR).
\end{acknowledgements}

\end{document}